\documentclass[a4paper,11pt]{article}
\usepackage{apacite,graphicx,array,multirow,longtable,nicefrac,url,setspace}

\newcommand{\keywords}[1]{\textbf{Keywords:}\quad #1}

\begin{document}

\fontfamily{ppl}\selectfont

\title{A review of mass concentrations of Bramblings \emph{Fringilla montifringilla}: implications for assessment of \newline large numbers of birds}

\author{Tomas Svensson}

\maketitle


\begin{abstract}
Mass concentrations of birds, or lack of such, is a phenomenon of great ecological and domestic significance. Apart from being and indicator for e.g. food availability, ecological change and population size, it is also a source of conflict between humans and birds. Moreover, massive gatherings or colonies of birds also get the attention of the public -- either as a spectacular phenomenon or as an unwelcome pest -- thereby forming the public perception of birds and their abundance. In the context of the mass concentration of bramblings (Fringilla montifringilla) in Sweden the winter 2019--2020, this work reviews the literature on this striking phenomenon. Winter roosts are found to amount to on the order of one million birds per hectare of roost area, but the spread between reports is significant. Support for roosts of up to around 15 million birds was found, but much larger numbers are frequently recited in the literature. It is argued that these larger numbers are the result of overestimation or, in some cases, even completely unfounded (potentially typos). While the difficulties related to the count of large numbers of birds can explain this state, it is unfortunate that "high numbers" sometimes displace proper numbers. Since incorrect data, and its persistence, may result in that incorrect conclusions are drawn from new observations, this matter deserves attention. As the Brambling is a well-studied species, the matter also raises concerns regarding numbers for mass concentrations of other species. It is recommended that very large numbers of birds should be recited and used with care: underlying data and methods of the original sources should be scrutinized. Analogously, reporters of large numbers of birds are advised to describe and document counting methods. In particular, number estimates based on flock volume and bird density should be avoided.
\end{abstract}


\pagebreak
\tableofcontents
\pagebreak

\section{Introduction}
The occurrence, or absence, of large flocks of birds is linked to factors such as population size, food abundance and food distribution and can therefore carry important information on the status of, and changes in, the environment \cite{Hemery1981_Thermographie,Moller2019_EcoEvol}. Mass concentrations of birds is also an important source of conflicts with humans. As an example, tens of millions of red-billed quelea \emph{Quelea quelea} are, due to their impact on farming, annually poisoned or blown up in Sub-Saharan Africa \cite{McWilliam2004_EnvCons}. Similarly, various blackbirds -- most notably the red-winged blackbird \emph{Agelaius phoeniceus} -- aggregate in huge flocks and are considered a major pest in North America \cite{Linz2017_Blackbirds}. The European starling \emph{Sturnus vulgaris}, one of only three bird species on the IUCN list of the world’s worst invasive species \cite{Lowe2000_IUCN}, is yet another interesting example. Starlings are considered a serious problem in many countries over several continents, causing, for example, damage on fruit and berry farming and spread of disease \cite{Feare1992_ProcVertPest,Homan2017_StarlingsUSA}.

The bird species mentioned above are examples of birds that are reported to gather in millions. Such mass concentrations of birds are rare, and it is not more than around 50 of the worlds bird species that are reported to reach seven-digit numbers (see e.g. \citeA{Moller2019_EcoEvol}). As it is well known that counting large numbers of birds is extremely difficult, and since it matters if a bird count is 100 000 or 1 000 000 (a factor ten difference in potential crop damage or population size matters), it is well worth looking closer into how mass concentrations are counted and accounted for.
This work takes a closer look at the brambling \emph{Fringilla montifringilla} and its mass appearances in Europe during beech mast years. The brambling is the species in Europe that appears in largest flocks and is thus an interesting case study for discussing number estimates. The work is the result of an attempt to put the mass concentration in southern Sweden during the 2019--2020 winter into an international and historical perspective. After a description of the 2019--2020 events, a literature review is presented and counting methods discussed. The focus is on number estimation and roost density. For other aspects of the fascinating phenomenon of communal roosting of bramblings, the reader is referred to the existing literature. There are numerous ambitious studies of the topic, covering everything from relation to food abundance and snow cover \cite{JenniNeuschultz1985_OrnitBeob,Jenni1987_OrnisScand}, origin of birds and ring recoveries \cite{Schifferli1953_OrnitBeob,Jenni1982_OrnitBeob,Kjellen1993_Anser,Browne2003_RingMigr}, roost microclimate and communal roosting aspects \cite{Jenni1991_OrnisScand,Jenni1993_Ibis,Khil2011_Limicola,Arizaga2012_RingMigr,Zabala2012_ActaOrni} and foraging patterns and energy needs \cite{Granvik1916_FaunaFlora,Hemery1975_CrAcadSc,Hemery1976_LeTerre,Francois1978_Falco,NardinBrauchle1979,Nardin1985_NosOiseaux,Jenni1987_Ardea,Kjellen1993_Anser,Khil2011_Limicola} to behaviour of raptors in the vicinity of roosts \cite{Jenni1993_Ibis,Khil2011_Limicola,Zuberogoitia2012_JRaptRes}.

\section{Brambling ecology and behaviour}
The brambling \emph{Fringilla montifringilla} is one of the most numerous birds in the world. It breeds dispersed in northern forests of Europe and Asia, with a breeding range stretching all the way from Norway to eastern Siberia. The global population, which seems to be under moderate decline, is estimated to 100--200 million pairs, while the European population is limited to around 15--25 million breeding pairs (Birdlife International 2020). The species is highly migratory, but irruptive (i.e. exhibiting significant variation in migration patterns, see e.g. \cite{Newton2006_IrruptiveMigr,Newton2012_JOrnit}). While it is said to mainly migrate during night \cite{Newton1972_Finches}, some 100 000 bramblings are annually counted in Falsterbo, while leaving Sweden in daylight (on average almost one million finches are counted, but chaffinch \emph{Fringilla coelebs} dominates and typically 5--25\% are bramblings, \citeA{Kjellen2019_FiSk2018} and Kjell\'{e}n 2020, personal communication). While few ornithologists appear to have heard bramblings during the night, data from lighthouse suggests that the brambling indeed is a nocturnal migrant. Between 1886 and 1939, in an impressive and persevering Danish study, 1 568 bramblings were killed by and collected at Danish lighthouses, to be compared with e.g. 532 chaffinches \emph{Fringilla coelebs} and 1 569 willow warblers \emph{Phylloscopus trochilus} \cite{Hansen1954_Fyrfall}. \citeA{Alerstam1993_BirdMigration} has also noted bramblings leaving Sweden, heading out over the Baltic sea, at dusk.

In essence, the European bramblings moves towards the southwest until it finds satisfactory food resources \cite{Jenni1987_OrnisScand}. Many end up in the vast beech forests of Central Europe, but the migration patterns vary significantly from year to year. In Sweden, an often mentioned example is an individual ringed when wintering in Halland (Southwest Sweden) in January 1965, just two winters after to be recovered in Caucasus \cite{BRC2020,Kjellen1993_Anser}. Other ringing recovery examples include a bird ringed in Blekinge (Southeast Sweden) in January 1986, the year after wintering in southwest France where it was found dead, and a bird ringed in Sm{\aa}land (South Sweden) in January 1955 later being killed in Spain in November the same year \cite{BRC2020}. Swiss ring recoveries shows that central Europe is also reached by bramblings with origin east of the Urals \cite{Jenni1982_OrnitBeob}. This nomadic character is considered to be a reason why the species, despite its wide breeding range, is monotypic.

\begin{figure}
  \centering
  \includegraphics[width=17cm]{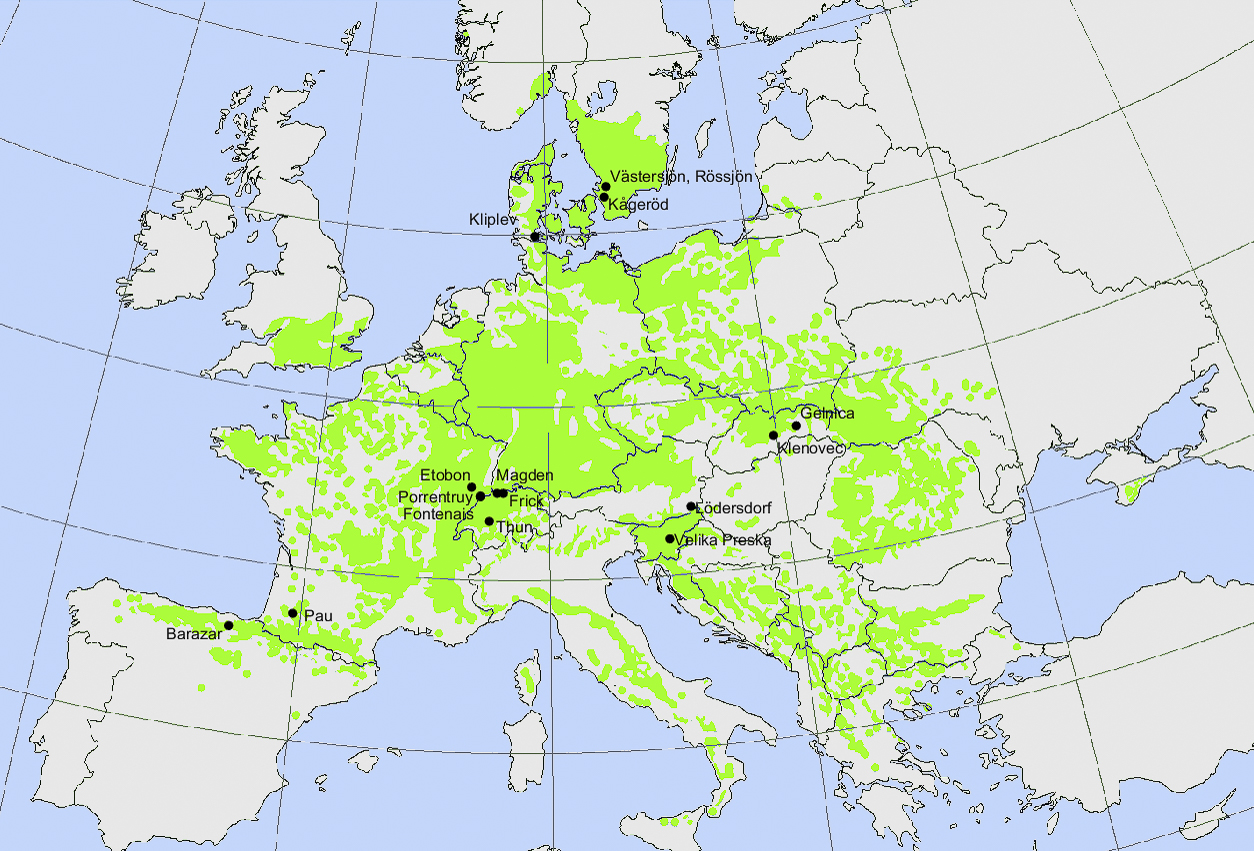}
  \caption{Distribution map of Beech \emph{Fagus sylvatica} in Europe, adapted from \protect\citeA{EUFORGEN2009}. As beech nuts is the primary winter food of bramblings, the map indicates where they can be expected to winter in large numbers. Winter roosts that has been described in detail, constituting the core of this study, are marked (cf. Table \ref{tab:roost_table}). Note that Oriental beech \emph{Fagus orientalis}, which can be found in e.g. Turkey and the Caucasus, is not included in this map.}\label{fig:BeechMap}
\end{figure}

When it comes to food, the brambling is an omnivore, although rather specialized. During breeding, bramblings have been found to rely on larvae of the Autumnal Moth (Epirrita autumnata), and its breeding success correlates with the strong cyclical fluctuations of this moth species \cite{Lindstrom2005_Oikos,Newton2007_MigrEcol}. During winters, a strong inclination for communal roosting (cf. \citeA{Beauchamp1999_BehavEcol} makes them responsible, albeit sporadically, for one of most spectacular mass concentrations of animals in Europe. Nuts of the European beech \emph{Fagus sylvatica} is their primary winter food and during beech mast years the abundance of food allows them to aggregate and form roosts of millions of birds (see e.g. \citeA{Jenni1987_OrnisScand}). The birds typically settle in coniferous forest sections in the vicinity of large beech forests. Interestingly, in a limited period of time 1960--1980, also corn fields in France played an important role, see \ref{subsec:corn_districts} and e.g. \citeA{Hemery1975_CrAcadSc} and \citeA{Dubois2008_BirdsFrance}. The same patches can be used for months if food supplies and snow coverage allow. The distribution of beech in Europe is shown in \ref{fig:BeechMap}, giving a rough indication of where European bramblings can be expected to winter in large numbers.

Historically, the appearance of millions of bramblings was often seen as a bad omen, as sign war, starvation or pest \cite{Holmgren1866,Granvik1916_FaunaFlora,Haikos1950_OrnitBeob}. Today, it is more of an attraction for nature lovers. However, it is worth keeping in mind that species for which a significant part of the population accumulates in limited areas are very vulnerable to, for example, hunting and poisoning. A contemporary example of this kind of vulnerability is the dramatic decline of the yellow-breasted bunting \emph{Emberiza aureola}. Once superabundant, rampant trapping has brought the population of this species down to around 10\% what is was only a few decades ago \cite{Kamp2015_ConsBiol}.

A single brambling requires some 25--30 kcal per day \cite{Hemery1975_CrAcadSc,Hemery1976_LeTerre,Kjellen1993_Anser}, corresponding to around 8 g of beech seeds (around 40 seeds, or around a fourth of their body weight). This is based on published energy values for tree-seeds: the nutritive material (seed excluding coat) of a seed carries 7 kcal per g dry weight and the dry weight of nutritive material per seed is around 0.12 g \cite{Grodzinski1970_TreeSeeds,Nilsson1979_Ibis}, resulting in around 0.84 kcal, or 3.5 kJ, per seed. Assuming 83\% energy utilization \cite{Kjellen1993_Anser}, this means that a single brambling requires some 40 seeds per day (27.5 kcal/day / (83\%$\cdot$0.84 kcal/seed) $\approx$ 40 seeds/day). Since the fresh weight of a whole seed is around 0.2 g, this corresponds to 8 g. These energy considerations indicate that one million bramblings consume on the order of 8 tons of beech seeds per day, summing up to around 1 000 tons for a four-month wintering stay. For a mast year, the beech nut production can be well above 1 ton/ha \cite{Kjellen1993_Anser,Oevergaard2007_FaktaSkog,Overgaard2010_PhDThesis}. Given than bramblings can fly up to at least 40 km from a winter roost \cite{Muehlethaler1952_OrnitBeob,Hemery1975a_CrAcadSc,Francois1978_Falco,Jenni1984_PhDThesis,Chalverat2003_SchwForst,Khil2011_Limicola}, the accessible area is around 500 000 ha. The fraction of beech forest in that area can thus even be relatively low (e.g. even below a 1\textperthousand level) and still support millions of bramblings for several months.

An interesting aspect of mass roosting is the massive amount of faeces deposited in the roost area. Not only does it produce a nasty odour, it also affects the ecosystem \cite{Chalverat2003_SchwForst}. In 2000, as an example, a new species of fungi was found in the aftermath of a huge brambling roost and was named \emph{Pseudombrophila stercofringilla} after the bird excrements \cite{Dougoud2001_MycoHelv}. Still, the impact of brambling roosts on forest ecosystems should be small compared to the how the enormous passenger pigeon \emph{Ectopistes migratorius} population -- a species also specialised on tree-seeds such as beech mast affected forest ecology in North America (see e.g. \citeA{Bucher1992_CurrentOrn} and \citeA{Ellsworth2003_ConsBiol}). In general, the knowledge of ecological functions of birds is far from complete \cite{Sekercioglu2006_TrendsEcoEvol} and the role bramblings have played in European beech forests, if any, is not clear.

\section{The mass concentration in Sweden 2019--2020}\label{sec:the_mass_conc_in_Sweden}
\begin{figure}
  \centering
  \includegraphics[width=17cm]{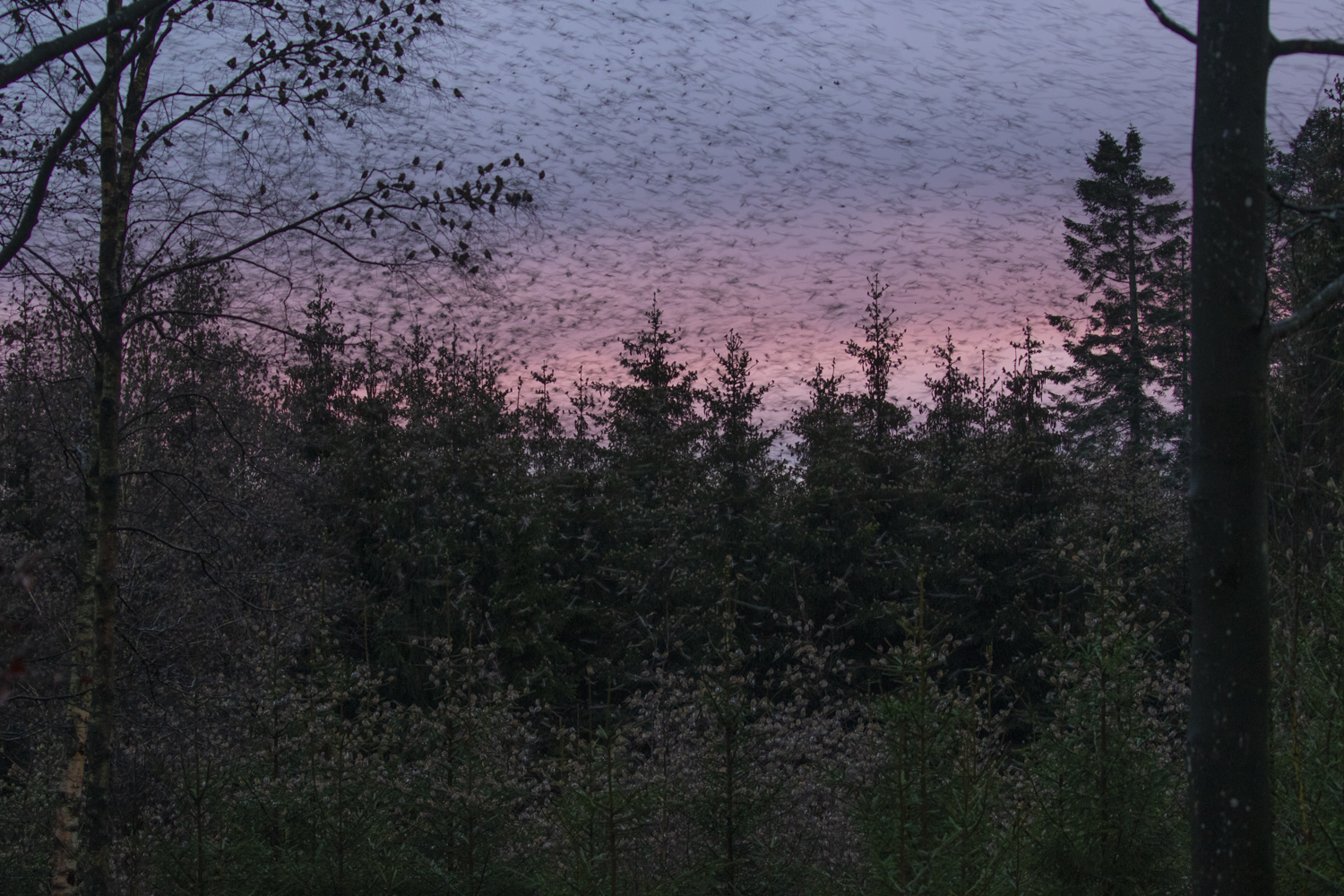}
  \caption{Evening at the roost at lake R\"{o}ssj\"{o}n, Sweden on 4 February 2020. View towards the northern side of the younger section.}\label{fig:EveningView}
\end{figure}

\begin{figure}
  \centering
  \includegraphics[width=17cm]{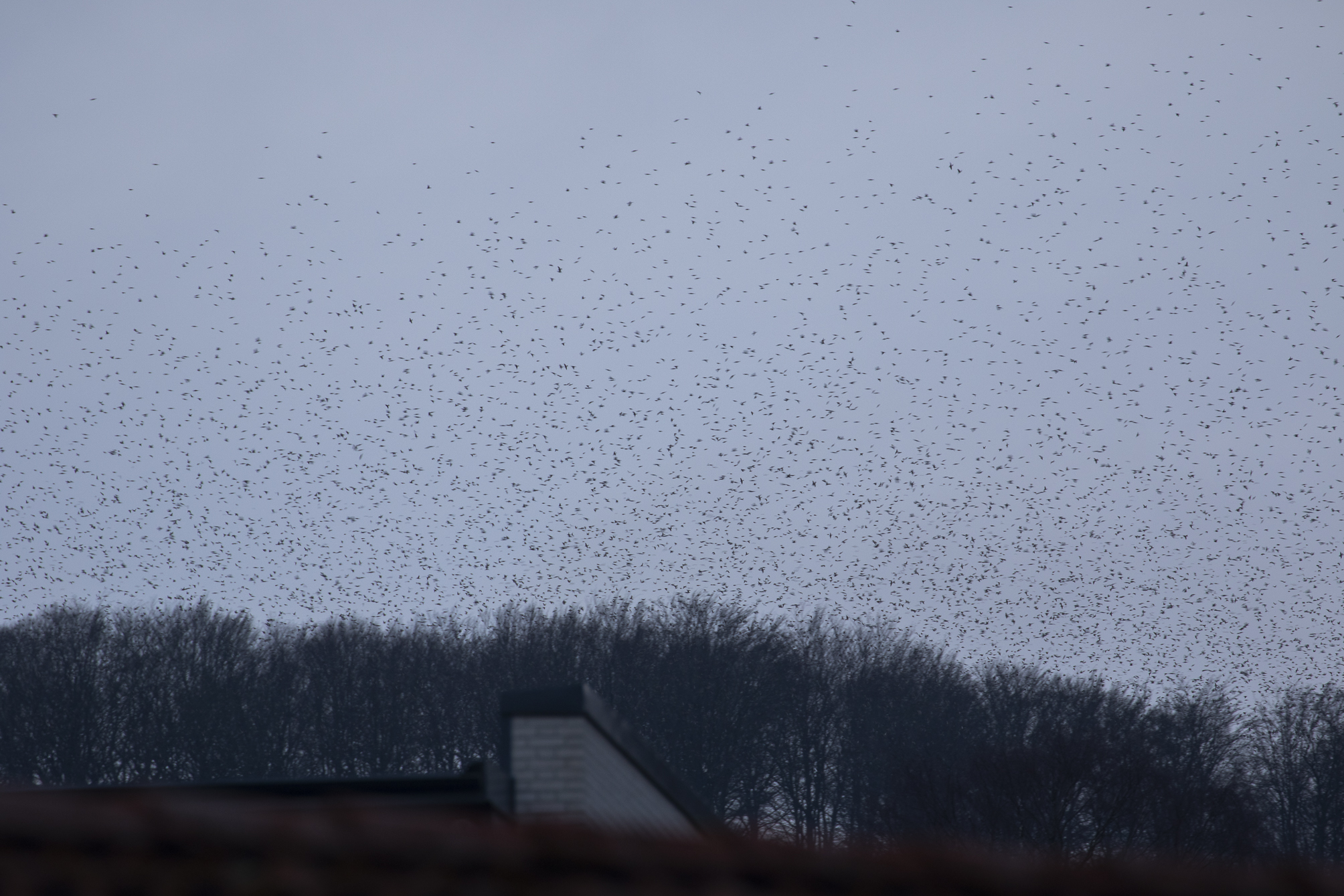}
  \caption{Distant stream of bramblings when passing B{\aa}stad 22 December 2020. The photo was taken with a Canon EOS7D with EF100--400 mm f/4.5--5.6L IS II USM at its maximum focal length, giving a 3.23$^\circ$ horizontal field-of-view. From the size of the birds (0.15 m physically, and around 15 pixels on the 5427$\times$3648 pixel sensor), this stream is estimated to be around $0.15/\tan(15\textrm{px}/5427\textrm{px}\cdot3.23^\circ)\approx$ 1 km away. By tedious manual marking of the individual birds, it is known that the photo contains around 8260 birds. The camera field-of-view at 1 km is around 56 m. Radar measurements has shown that bramblings fly at 15 m/s \protect\cite{Alerstam2007_PLoS} and after accounting for around 2 m/s head wind (as reported by the closest wind station), the 8 260 birds should pass the field of view in around $56/13\approx 4.3$ s. This corresponds to an intensity of around 1 900 birds per second.}\label{fig:Bastad_December}
\end{figure}

After an extremely warm and dry summer 2018 and good conditions for beech flowering during 2019, the beech seed crop of 2019 turned out to be enormous. Although the quantitative beech mast counts, operated by Swedish University of Agricultural Sciences, unfortunately was discontinued a few years ago, there is no doubt that 2019 was an extreme mast year. Rough counting, by the author, indicated that a single large beech tree could carry on the order of 100 000 shells (cupule), corresponding to 200 000 seeds. Assuming that one hectare of beech forest is equivalent of 100 large trees, the seed production could be on the order of 20 million per hectare (i.e. around 4 tons/ha). As a comparison, highest local beech mast production during the beech mast counting 1989--2006 amounted to 14 million/ha \cite{Oevergaard2007_FaktaSkog,Oevergaard2007_Forestry,Overgaard2010_PhDThesis}. Furthermore, Prof. Sven Nilsson has conducted semi-quantitative monitoring of beech mast in Sweden since 1971, ranking the crop from 0 (no crop) to 5 (massive crop), and reports that there has not been a year in this series with as much beech mast as 2019 (Nilsson 2020, personal communication).  Without standardized quantitative data, it is however difficult to know exactly how the 2019 mast year compares to other mast years. In addition to the rich beech crop, it is also interesting to note that the brambling seems to have had a good breeding season in Sweden. The LUVRE project (www.luvre.lu.se), monitoring birds in the northern birch forests since the 1960s, reports their highest number of ringed juvenile bramblings since the ringing start in 1983, and the third highest number in terms per juveniles per adult ({\AA}ke Lindstr\"{o}m 2020, personal communication).
That masses of Bramblings this year ended their migration in southern Sweden, taking advantage of the food abundance, became clear in November--December. Large flocks were reported from various areas of Scania (Sk{\aa}ne). A massive afternoon movement over the city of B{\aa}stad on December 22, witnessed by the author and Stefan Svensson (cf. Fig. \ref{fig:Bastad_December}), involved millions of birds and indicated that a roost was to be found in the area. The roost was found by the author on January 4 \cite{Svensson2020_Anser} and was located on the south-facing slopes of the Hallands{\aa}s Horst, just north of lake R\"{o}ssj\"{o}n. Photos and moviews can be found at \url{https://www.flickr.com/photos/tomas_s/albums/72157712566702823}. The roost area was estimated to 5.6 ha, via night time sounds in combination with visual inspection of excrement layers and their boundaries (droppings) together with areal analysis using map tools supplied by Lantm{\"a}teriet (www.lantmateriet.se). The roost consisted of, respectively, approximately 2.6 ha of 31 year old and 3.0 ha of 39 year old plantations of European spruce \emph{Picea abies}. The younger parts were dense with only around 2 m between tress while the older parts where thinned out and had 4--5 m distance between trees. Tree height ranged from 15 to 25 meters. Fig. \ref{fig:EveningView} shows a view from the roost perimeter

Employing a simple square grid model on these tree spacings, and reducing the count by 10\% to account for small opening and forest roads, this results in the following rough tree count estimate:
\begin{equation}\label{eq:tree_estimate}
  N_{trees}=0.9\cdot(2.6\cdot(100/2)^2+3.0\cdot(100/4.5)^2 )\, \approx 7\,0
  00\,\textrm{trees}
\end{equation}

\begin{figure}
  \centering
  \includegraphics[width=17cm]{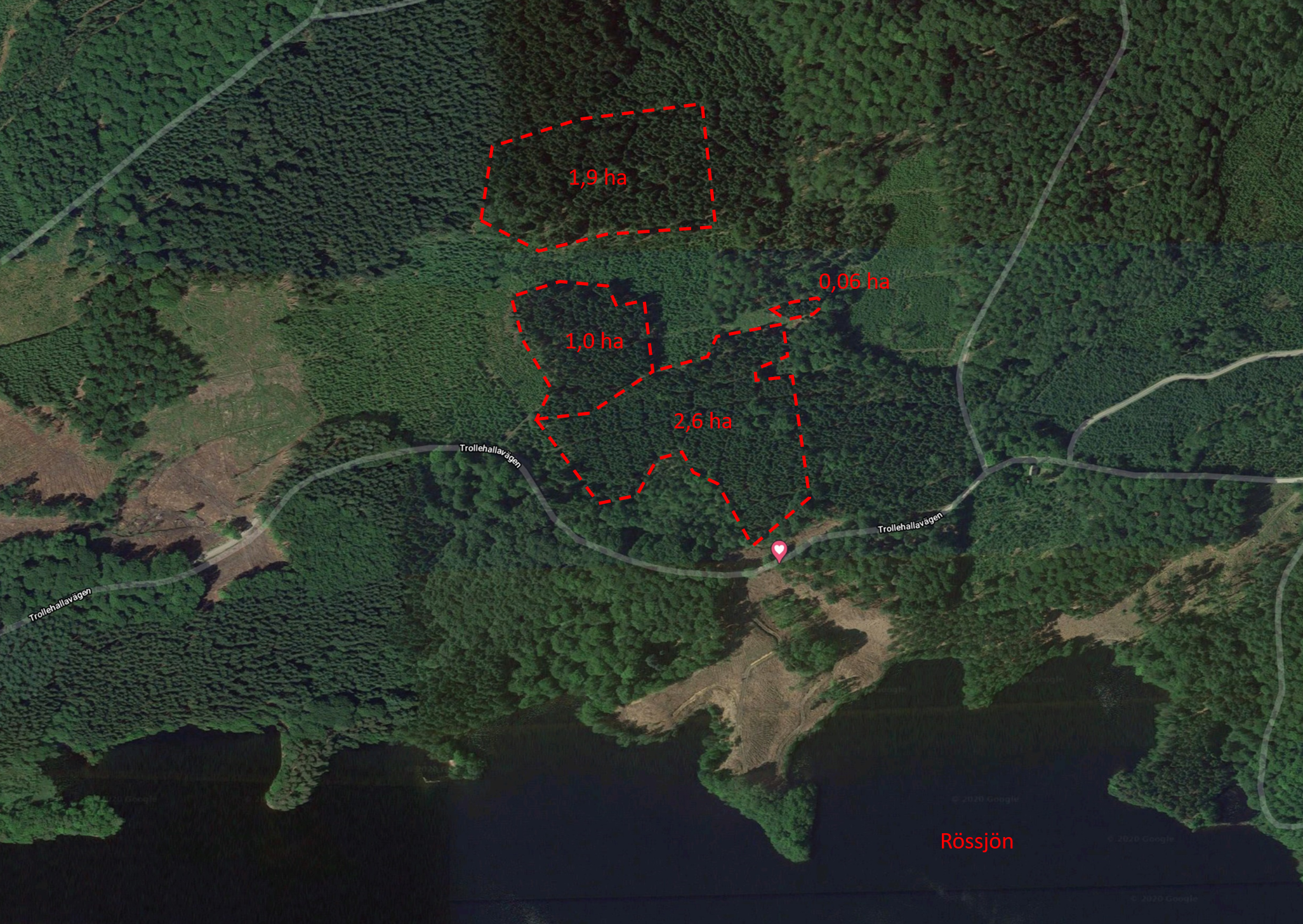}
  \caption{Annotated map showing the 5.6 ha roost area that appeared to be used simultaneously in early January 2020 (from Google Maps). The area was demarcated based on droppings and evening sound.}\label{fig:Roost_area}
\end{figure}

A map of the roost area is shown in \ref{fig:Roost_area}.

On-site counting at the 2019--2020 site was very difficult, as it was impossible to get a good overview of the roost. Monitoring morning lifts or evening fly-in, was thus not easily done. The best number estimate is most likely that from B{\aa}stad on December 22, assuming that all of the birds passing there was headed for the same roost and that most of the roost had the same whereabouts this day.  Despite being witnessed by rather experienced migration counters, the flow over B{\aa}stad was overwhelming. From analysis of photos, the front of Bramblings was over one kilometre wide at its peak intensity (determined via photographs).

\begin{figure}
  \centering
  \includegraphics[width=17cm]{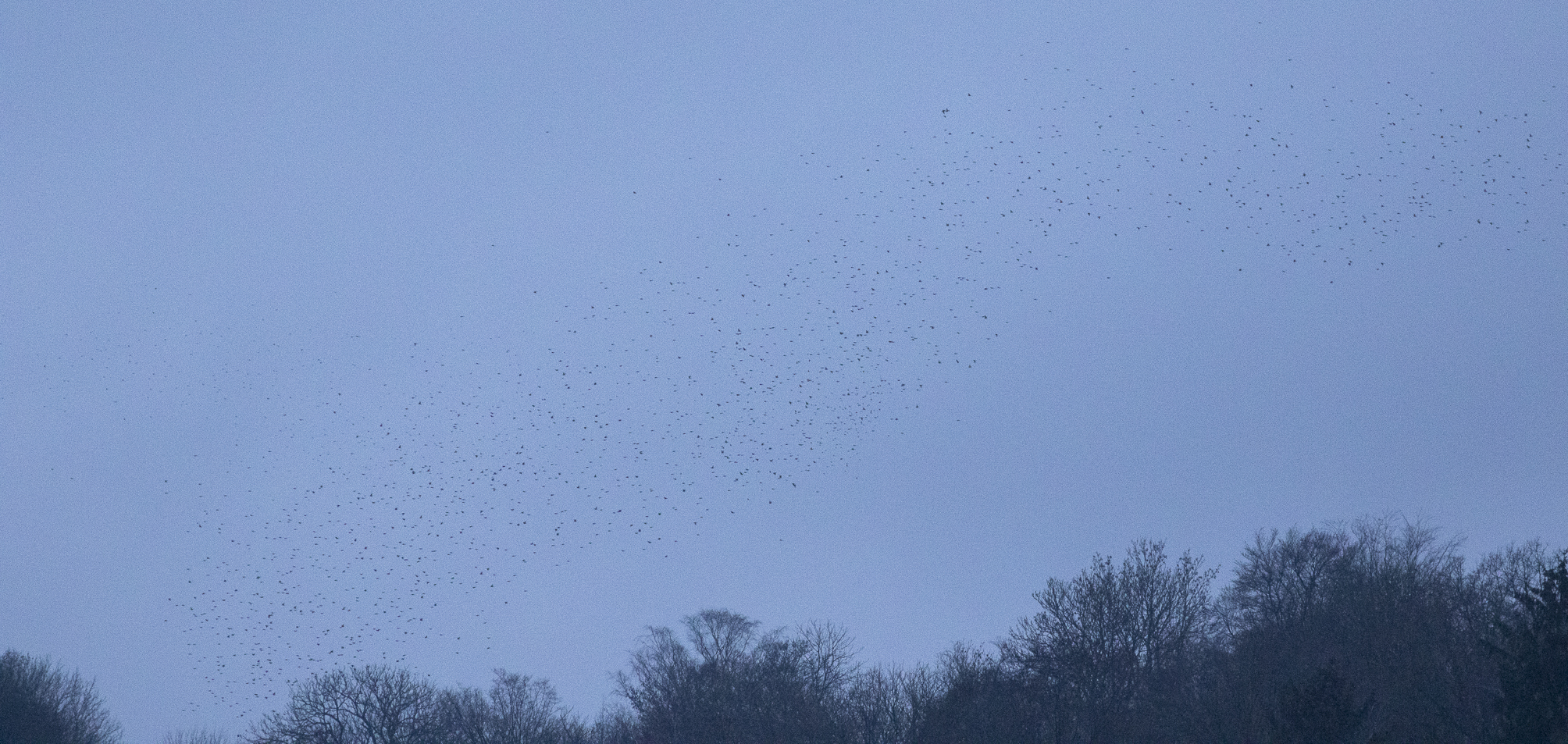}
  \caption{Photo taken at a time where the movement was considered to be nearly dead (B{\aa}stad 22 December 2020). The photo was taken 15:33, but the intensity increased again after this dip. The movement ended at around 15:40). Excluding a faint stream in the background, the photo contains around 1575 birds (manually marked) flying at a distance of around 1.2 km. The corresponding instantaneous intensity is around 300 birds per second.}\label{fig:Bastad_December2}
\end{figure}

Thousands of bramblings was in the binocular field of view and passed within a matter of seconds. A number estimate was thus reached only after analysis based on timing and photographs. Our field notes state that the movement occurred during around 5 minutes around 13:40, and then during 45 minutes from 14:55 to 15:40 (it was almost dark in the end, so the Bramblings seem to have been late this day -- the sunset was at 15:33). Analysis of photo of a distant streak (where birds are caught from the side, simplifying analysis) showed that 1 900 birds passed per second (1 900 s$^{-1}$, see Fig. \ref{fig:Bastad_December} and its caption for details). At the peak intensity we believe that the total intensity should be two to three times of that. An analysis of a photo taken at a time where we considered the migration as "dead", revealed an intensity of 300 s$^{-1}$ (see Fig. \ref{fig:Bastad_December2}).

\begin{figure}
  \centering
  \includegraphics[width=12cm]{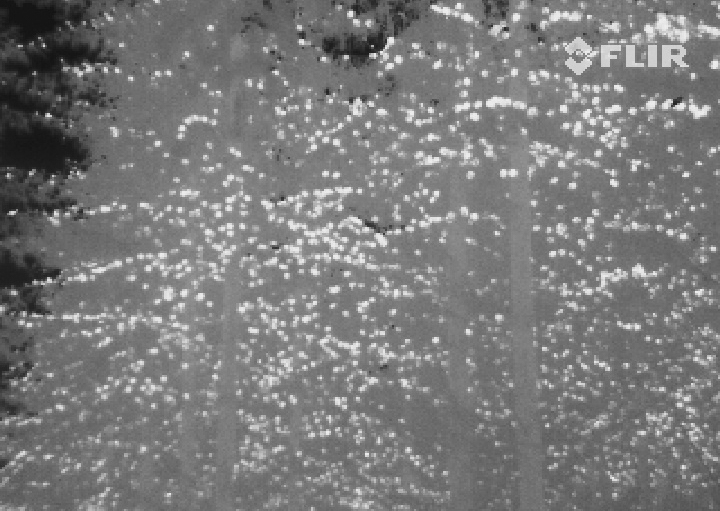}
  \caption{Thermal photograph taken with a hand-held FLIR Scout TS24 Pro. The photo was taken just outside the roost perimeter in the dark. It was impossible to walk further without causing chaos. The slightest sound or light increased the chattering and made many birds take off with the sound of storm winds.}\label{fig:IR_photo}
\end{figure}

The numbers are breath-taking. In Falsterbo, with its annual standardised migration counts, finches are typically counted in a few thousands per minute on a good day. We (the author and Stefan Svensson) judged that the most intense period lasted around 20 minutes, and that the total amount of birds could be calculated from 30 minutes with 1000 s$^{-1}$ and 20 minutes with 5 000 s$^{-1}$, resulting in 7.8 million birds. As a lower bound, and since field notes are imperfect and we did not take photos systematically throughout the Brambling passage, we propose to use 40 minutes as to total duration and restricting the intense period to 10 minutes and 4000 s$^{-1}$. This approximate lower bound amounts to 4.2 million birds. As an upper bound, maybe 30 minutes with 1500 s$^{-1}$ and 20 minutes with 6 000 s$^{-1}$ is reasonable, summing up to 9.9 million.  A reasonable, albeit rough, range for the number of Bramblings involved in this movement is thus 4--10 million. As a comparison, when on-site on January 4th, the author together with Nils Kjell\'{e}n and Ola Ellestr\"om very roughly estimated that we witnessed, visible from our side of the roost area, a fly-in intensity of around 2000 s$^{-1}$ during 40 minutes, summing up to around 5 million birds. The sound and visual experience was stunning.
Some efforts were also made to estimate the number of birds roosting per tree, but no firm conclusion could be reached. Infrared photographs indicate that ten birds easily can sit on a single twig during the night (see Fig. \ref{fig:IR_photo}), and observations and photographs taken at dusk indicate that on the order of 1 000 birds can fit in a single spruce tree. Still, the number of birds per tree during the night remains an open question and more efforts are needed to elucidate this matter. As movements seems occur after dark, and since the bramblings are disturbed when approached, installation of infrared cameras in the roost may be an interesting approach.

Although bramblings winter in Sweden regularly, few Brambling winter roosts have been found. Prior to the roost described above, only three roosts have been located and described in detail \cite{Granvik1916_FaunaFlora,Mathiasson1960_SvNat,Kjellen1993_Anser}. One of these roosts was found 1993 very close to the location described above: just north of the neighboring lake V{\"a}stersj{\"o}n, on the same south-facing slopes \cite{Kjellen1993_Anser}. Another roost was found was found some 36 km south from these locations during the 1915--1916 winter \cite{Granvik1916_FaunaFlora}. This part of Scania has vast beech forests and is relatively sparsely populated. It is therefore likely that even rather large roost can be overlooked. As an example, during the research for this article, the author stumbled across a YouTube-video filmed by fishermen in February 2012 showing enormous flocks of bramblings flying over lake V\"astersj\"on \cite{Hafstrom2012_YouTube}. No roost was found this year. There are also years when ornithologists have reported million level flocks in various parts in southern Sweden, still without any roosts having been located.

Given the climate change that is upon us, it can be expected that bramblings will winter in southern Sweden in increasing numbers. The winter 2019--2020 was in fact, according to standard meteorological definitions, not a winter season at all and the average temperature was four degrees above normal \cite{SMHI2020}. Scania is rarely covered in snow nowadays and the frequency of beech mast years have increased in recent years \cite{Oevergaard2007_Forestry}, although long-term variability also must be considered \cite{Drobyshev2014_AgriForMet}.

\section{A review of mass concentration reports}

It is sometimes argued that the brambling is the bird that gathers in the largest numbers of all (Newton 1998). A better guess is perhaps the red-billed quelea \emph{Quelea quelea} which is reported to occur in numbers up to 100 million \cite{Hancock2015_BirdsBotswana}. There are also several other bird species are potential “records holders” (see e.g. \citeA{Moller2019_EcoEvol}), but given the difficulties in bird counting, it appears very difficult to get such a question settled. If including extinct species, the passenger pigeon \emph{Ectopistes migratorius} will make settlement easier: the species has even been claimed to occur in billion level flocks (a sad but important reminder that what is common today may be go extinct more rapidly than we expect, cf. \citeA{Murray2017_Science}).

Turning back to bramblings, gatherings of millions are far from being reported annually. This is expected, since very large roosts appear to form only during beech mast years. Throughout the years, however, numerous accounts of the phenomenon are available thanks to authors from several European countries. Unfortunately, many reports neither provide details on how numbers were estimated, nor details regarding roost area. Many authors express how difficult it was to count properly and there is in many cases massive disagreements regarding the actual number of birds involved. This review aims at elucidating this matter and the main focus is therefore on reports that gives details on number estimations and roost area.

\subsection{Approaches for number estimation}
Three different approaches to number estimation was encountered during this work:

\begin{itemize}
    \item[A.] Stream intensity and time, $N = I\cdot T$
    \item[B.] Flock volume and density, $N = B\cdot H\cdot L\cdot \rho = B\cdot H\cdot v\cdot T\cdot \rho$
    \item[C.] Tree count and birds per tree, $N = N_{\textrm{trees}} * N_{\textrm{per\quad tree}}$
\end{itemize}

In method A, the number of birds $N$ is reached by estimating the intensity of birds, $I$, passing an observer (birds/s) in combination with time duration $T$ of the passage of that intensity (since intensity will fluctuate, new estimates should ideally be done continuously). The intensity can either be estimated directly (instantaneously) by the observer, or it can be done afterwards via analysis of photographs. Method B, on the other hand, focusses on flock volume (width $W$, height $H$ and length $L$) and bird density $\rho$ (birds/m$^{3}$), where flock length is reached by assuming a certain ground speed $v$. Many works assume bramblings fly at 60 km/h $\approx$ 16.7 m/s, while radar measurements indicate that their ground speed is 15 m/s $=$ 54 km/h \cite{Alerstam2007_PLoS}. This inherent overestimation of 10\% should of course be avoided, even though it in most cases will be negligible compared to other uncertainties. Method C is very different: instead of counting flying birds, the focus is on roost parameters. The number is reached by multiplying the number of trees in the roost by the average number of birds per tree.

Making any general statement on the accuracy of the different methods, and how they compare, is difficult since it depends on how the individual parameters are estimated. However, as various studies are discussed in subsequent sections, it will be argued that choosing to go via flock volume and density should be avoided. In many works, width and height in method B is set without proper motivation (keep in mind that difficulty of estimating distances by eye is widely recognised). In addition, there is far from any agreement on the average bird density in streams of bramblings. In fact, it should be expected that this varies significantly from roost to roost, and even from day to day at a single roost (depending on e.g. weather and foraging patterns). This means that method B is most likely far more error-prone than other methods. If width and height is estimated from photos (from size variations of birds in the picture, see Fig. \ref{fig:density} for an example), method B can become more accurate, albeit at the same time essentially turning into a complicated version of method A. 

In the future, new approaches may help in reaching better number estimates. Airborne thermography was proposed by already 40 years ago  \cite{Hemery1981_Thermographie}, and with the increasing availability of infrared and thermal cameras (or even normal video cameras) roosts will likely be studied in more detail in the future. Approaches based on quantitative monitoring of sound or droppings is also conceivable. In addition, increasing availability of counting software, for example based on artificial intelligence, will simplify counting based on films or photographs. 

\subsection{Reports with details on number and roost area}
\begin{figure}
  \centering
  \includegraphics[width=17cm]{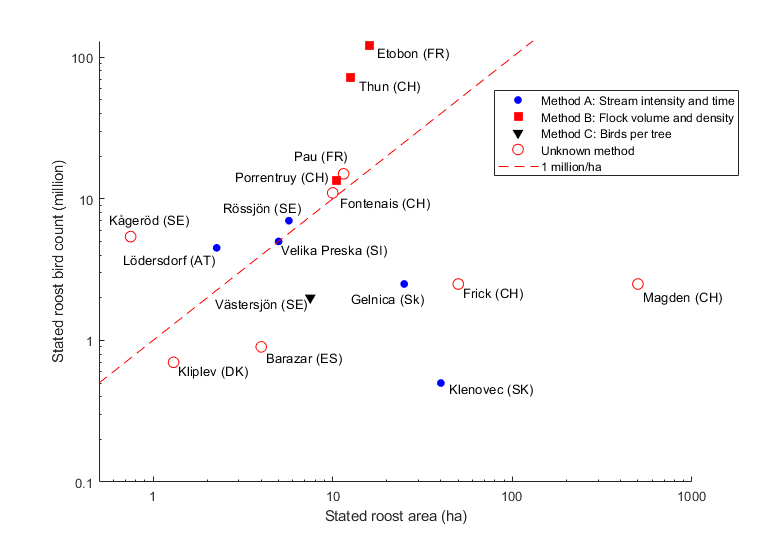}
  \caption{Scatterplot in logarithmic scale showing the variability in the data from Table \ref{tab:roost_table}. The four datapoints with roost area above 20 ha (Gelnica, Klenovec, Frick and Magden) are, in the author's opinion, obivous cases where the roost area has been significantly overestimated. Similarly, the number estimates for the roosts in K{\aa}ger{\"o}d, Thun and Etobon lack solid motivation and is in the present work argued to be the result of overestimations.}\label{fig:scatterplot}
\end{figure}

Reports that provide details on number and roost areas are listed and briefly summarised and commented a in Table \ref{tab:roost_table} . The spread in reported numbers and roost areas are significant, and both number estimation and roost area definitions seem to vary greatly (cf. Fig. \ref{fig:scatterplot}). When disregarding works that seem to have a too wide definition of roost area, the roost density (birds per area) varies from some 0.3 million/ha all the way up to 7.5 million/ha. This corresponds to a factor 25 between lowest and highest estimate of bird per roost area, a spread that motivated further scrutiny. It turns out to be very difficult to condense this material into any simple rule of thumb, but based on careful reading of the material, I come to the conclusion that the number of birds per roost area is on the order of one million/ha. There are a few works where much larger roost densities are reported, but a closer look reveals the underlying numbers lack adequate motivation. There are also some reports where densities are much lower \cite{Fulin2017_Tichodroma,Kestenholz1993_OrnitBeob}, but this is likely related to overestimation of actual roost area used during the night (see comments in Table \ref{tab:roost_table} ). In the end, however, it should be expected that the density per roost area will vary significantly with, for example, age, size and density of trees, temperature, local variation of microclimate, total suitable area in relation to the number of bramblings accumulated during the winter etc. As an example, the large Slovenian roost in 2018--2019 comprised large, older trees (Miheli{\v{c}} 2020, personal communication) while the Swiss roost in 1946--1947 was situation in an area with young trees, most below 8 m in height \cite{Gueniat1948_OrnitBeob}. It should not be expected that the birds roost will have similar densities when the vegetation can differ that much.

\pagebreak

\begin{center}
\singlespacing
\begin{longtable}{p{2.5cm} p{1.5cm} p{1cm} p{6.7cm} p{3.1cm} }
\caption{List of studies that provide estimates of both number of birds and roost area. It should be noted that it roost area is most likely defined and estimated in very different manners. Studies that include a roost map are marked with an asterix (*), and a second asterix is added if the work also explicitly states that this area was demarcated based on excrement layers. The spread in estimates of number of birds and roost area is visualized in Fig. \ref{fig:scatterplot}}\label{tab:roost_table} \\
\hline
Location \newline Year & Roost area & Count method & Number estimate and remarks & Reference \\
\hline
K{\aa}ger{\"o}d \newline Sk{\aa}ne \newline Sweden \newline 1915--1916 & 0.5--1 ha & ? & 5.4 million, based on 45 min fly-in at a presumed intensity of 2000 s$^{-1}$ (without further motivation). The roost area is stated to be \emph{n{\aa}got tunnland} in Swedish, i.e. around one barrel of land (0.5 ha). & Granvik, 1916 \\

Porrentruy \newline Ajoie \newline Switzerland \newline 1946--1947 & 10.5 ha** & B & 11--16 million, based on 45 min morning lift-off in a stream 100 m wide and 5--7 m high, a ground speed of 60 km/h and 0.512 birds/m$^3$ (corresponding to an average intensity of 4000--6000 s$^{-1}$ ). & Gu{\'e}niat, 1948 \\

H{\"u}nibach \newline Thun \newline Switzerland \newline 1950--1951 & 12.6 ha* \newline (2$\times$6.8) & B & 72 million, based on 45 min fly-in in a 200 m wide and 4 m high stream, a ground speed of 60 km/h and 1 bird/m$^3$, thereafter doubled since there were two similar adjacent roost areas (corresponding to an intensity of 13 300 s$^{-1}$ per roost). Other assessors propose far lower numbers \shortcite{JenniNeuschultz1985_OrnitBeob}. & M{\"u}hlethaler, 1952 \newline Schifferli, 1953\\

Pau \newline Pyr{\'e}n{\'e}es-Atl. \newline France \newline 1964--1965 & 10--13 ha & ? & 15 million stated without further details. & Alberny, 1965\\

Etobon \newline Haute-Sa\^{o}ne \newline France \newline 1977--1978 & 16.1 ha & B (A) & 121 or 12 million, based on two different calculations. The lower number is based on an analysis of a photo with 5 565 birds over a width of 32 m in combination with 70 min fly-in at 60km/h (i.e. 2 900 s$^{-1}$). The larger (unreasonable) number, which is the one stated in the conclusion, is reached after arguing that the stream was 250 m wide and 50 m high and contained 0.138 birds/m$^{3}$ (i.e. 29 000 s$^{-1}$). & Nardin, 1979 \\

Magden \newline Switzerland \newline 1990--1991 & 5 km$^2$ \newline (500 ha) & ? & 2--3 million, stated without any further detail. Compared to other roosts, the area appears to be unreasonably large. & Kestenholz, 1993 \\

Frick \newline Switzerland \newline 1992--1993 & 0.5 km$^2$ \newline (50 ha) & ? & 2--3 million, stated without any further detail. Compared to other roosts, the area appears to be unreasonably large. & Kestenholz, 1993 \\

V\"{a}stersj\"{o}n \newline Sk{\aa}ne \newline Sweden 1992--1993 & 7.5 ha & C & 2 million, stated as a minimum based on around 7 500 trees and 200--500 birds per tree. Large number of dead and dying birds at the site. A similar number was reached in a fly-in count \shortcite{Lithner1995_FaglarNV}. & Kjell{\'e}n, 1993 \\

Fontenais \newline Ajoie \newline Switzerland \newline 2001--2002 &	10 ha** & ? & 10--12 million. Not exactly clear how this span was inferred. The estimation was based on a 32 min fly-in of a stream estimated to be 50 m wide, 15 m high and moving at 60 km/h. An intensity of 5000 s$^{-1}$ is stated as a minimum. The given 10--12 million range would correspond to calculation on the stream estimate and using 0.4--0.5 birds/m$^3$. Area description and map not in perfect agreement. & Chalverat, 2003 \\

Gelnica \newline Slovakia \newline 2008--2009 & 25--30 ha & A & 1.5--3.5 million, based on photo analysis. Roost area not given in the article but from Fulin (2020) in personal communication. Compared to other roosts, this estimate seems unreasonably large (but Fulin also mentioned that the area included several hectares of lake). & Fulin, 2017 \newline Fulin, 2020 \\

L\"{o}dersdorf \newline Austria \newline 2008--2009 & 2.26 ha & A & 4--5 million, based on photo analysis on several locations around the roost. & Khil, 2011 \\

Barazar \newline Basque \newline Spain \newline 2010--2011 & 3--5 ha & ? & 0.9 million. Droppings very unevenly distributed in the 50 ha large area mention in the article. Actual roost much smaller, and limited to patches of Lawson cypress (\emph{Chamaecyparis lawsoniana}), but not discussed explicity. In personal communication, Zabala (2020) guessed that perhaps 3--5 ha may have been used for actual roosting and that the number may have been in the span 0.7--12 million birds. & Zabala, 2012 \newline Zabala, 2020 \\

Klenovec \newline Slovakia \newline 2016--2017 & 40 ha & A & 0.5 million, based on photo analysis. Compared to other roost, the area seems unreasonably large. & Fulin, 2017 \\

Velika Preska \newline Slovenia \newline 2018--2019 & 5 ha & A & 5 million, based on photo analysis. Series of photos were taken during morning lift-offs (ground speed estimated by measuring how long it took for birds to pass the camera field of view, done by following individual birds and marking start and stop times using a voice recording). In personal communication Miheli{\v{c}} (2020) states that the roost area may have been up to around 9 ha. & \shortciteNP{Tout2019_RareBirdAlert} \newline Miheli{\v{c}}, 2020 \\

Kliplev \newline Denmark \newline 2019--2020 & 1.3 ha**	& ? & 0.2--1.2 millions. Various number estimates from several observers, typically based on estimates of volume and bird density of flocks flying over the roost area. Roost area well covered by excrements measured to 1.3 ha on February 2 (Hansen, DOF-basen 2020), and in the Facebook forum “Feltornitologen” there is a map showing marking a roost area comprising of 2--2.5 ha (Martinek 2020). & \shortciteNP{DOFbasen2020_Bergfink} \newline Feltornitologen, 2020 \\

R\"{o}ssj\"{o}n \newline Sk{\aa}ne \newline Sweden \newline 2019--2020 & 5.7 ha** & A & 4--10 million based on photo analysis from a fly-in around 20 km from the roost. More details in Section \ref{sec:the_mass_conc_in_Sweden}.	& This study, 2020 \\

\hline
\end{longtable}
\end{center}

\subsection{Regarding the alleged 70 million roost in Thun 1950--1951}
\begin{figure}
  \centering
  \includegraphics[width=17cm]{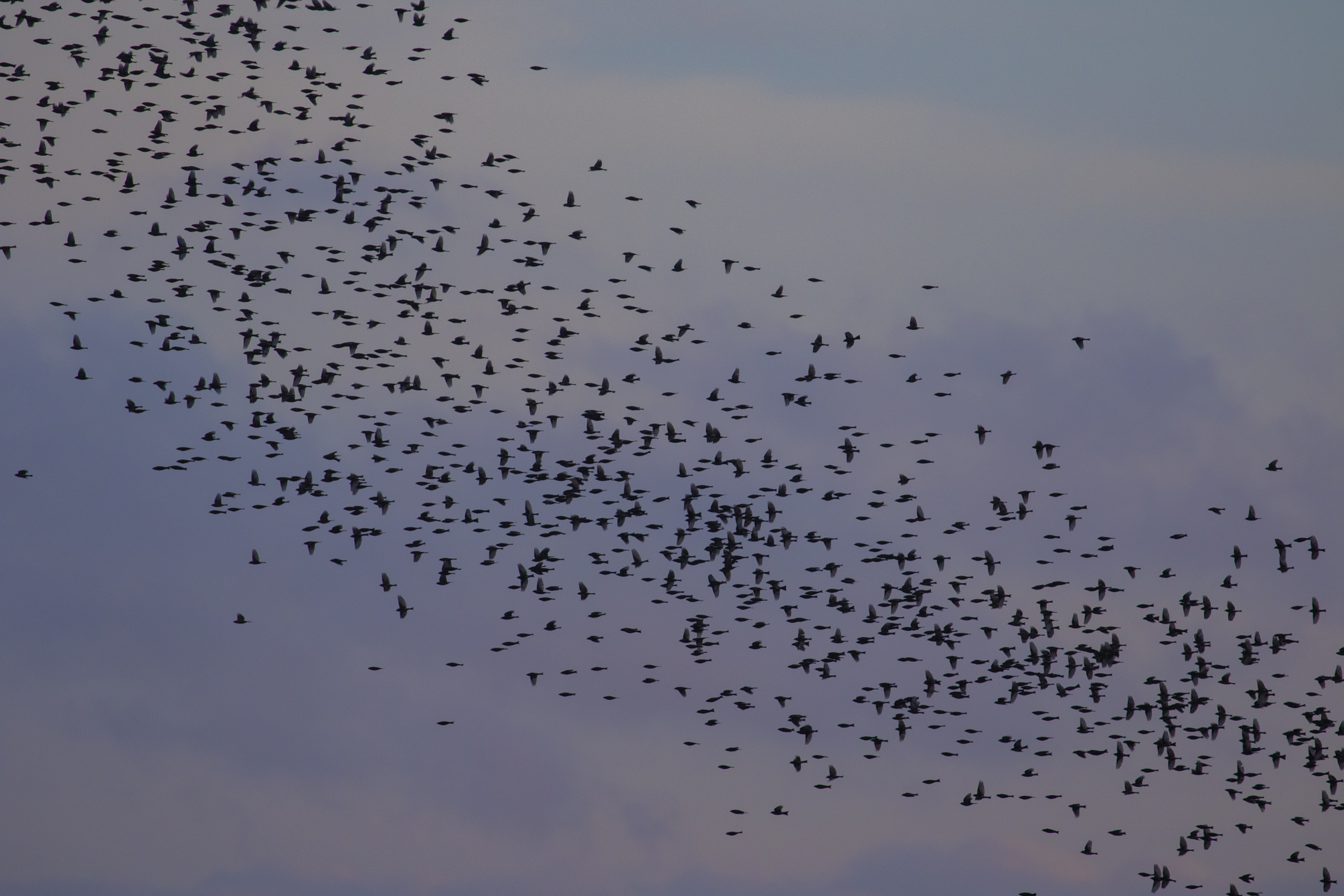}
  \caption{Flock density analysis example. The photo contains around 1 217 bramblings (manually marked) and is taken with a 3.23$^\circ$ field-of-view lens (Canon EOS7D with EF100--400 mm f/4.5--5.6L IS II USM at its maximum focal length). Bramblings are around 15 cm long and occupy in the range of 68--80 pixels, which corresponds to a distance range of 182--214 m. Since we have neglected length variations, this range should serve as an upper limit of the flock width (assuming a 10\% size difference from the smallest to the largest bird, one could argue that the flock width could be as low as around 10--15 m). The stream is around 3 m high, and the length captured by the photo is around 11.2 m. An lower limit for the bird density $\rho$ is reached when using the upper width estimate: $\rho_{min}=1217 / (3\times32\times 11.2 \approx 1.1\,\textrm{m}^{-3})$. With size variations in mind, the density could be as high as 3 m$^{-3}$. While this shows that bramblings can fly much denser than e.g. 0.1 m$^{-3}$, it should be noted that they often flew much less dense than this. In fact, the photo shown was taken when the bramblings flew particularly dense (at the roost site, 14 January). For reference, the flock length captured here is around 11 m. Assuming that the birds fly at 15 m/s \protect\cite{Alerstam1993_BirdMigration}, this stream corresponds to an intensity of around 1 600 birds per second.}\label{fig:density}
\end{figure}

The famous and frequently recited 72 million roost from Thun, Switzerland 1950--1951 \cite{Muehlethaler1952_OrnitBeob}, often recited as 70 million, deserves a closer look. The number comes from a doubling of an estimate of birds arriving to one of two adjacent roost areas separated by around 300 m. The stream of birds was estimated to be 200 m wide and 4 m high, and the duration of the fly-in was 45 min. Assuming a ground speed of 60 km/h, the length of the stream was estimated to 45 000 m. M\"{u}hlethaler believes that 1 bird/m$^3$ (1 m$^{-3}$) is a reasonable lower limit for the bird density and thus reach 200 m $\cdot$ 4 m $\cdot$ 45 000 $\cdot$ 1 birds/m$^3$ = 36 million birds. This corresponds to an average intensity of 13 000 birds per second, per roost. Along with some 30 million birds in other parts of Switzerland this winter, this roughly corresponds to the complete European post-breeding population. Although this number estimate received some support from \citeA{Schifferli1953_OrnitBeob}, it has been rejected by others. Lukas Jenni, most likely the ornithologist who has spent most time studying the winter habits of bramblings in Europe, argues that the estimate even may be more than a factor ten too high \cite{Jenni1984_PhDThesis,JenniNeuschultz1985_OrnitBeob}. The main objection is that the bird density is overestimated, and that it is very difficult to accurately determine width and height of a bird stream. While M\"{u}hlethaler suggested that a flock density of 1 m$^{-3}$ should be an underestimation, Jenni argues that 0.04--0.1 m$^{-3}$ is more reasonable which would bring down the number from 72 million to 2.8--7 million. Jenni refers to that stereo photography on brambling flocks during autumn migration exhibited density in the 0.05--0.7 m$^{-3}$ range, and that \citeA{NardinBrauchle1979} in their analysis of photographs estimated density to around 0.1 m$^{-3}$ in dense regions, and 0.04 m$^{-3}$ on average. The author of the present work, however, finds the analysis of Nardin \& Brauchle rather confusing and suspect that the range 0.04--0.1 m-3 underestimates how dense bramblings can fly at the roost. Jenni recites that measurements of density during autumn migration resulted in values up to 0.7 m$^{-3}$, and my personal experience is that the bramblings sometimes flew denser in connection to the winter roost than what the do, for example, during autumn migration in Falsterbo (Sweden). To evaluate this further, I evaluated two photos from occasions at the Swedish 2019--2020 roost where I found the bramblings to fly particularly dense. In these photos, the width of the stream was estimated from the variation in size (in pixels) of the photographed birds. The two photos were rather different in character (one photo of a smaller and well-defined flock, and one capturing part of a longer stream). The resulting density estimate was in both cases 1--3 m$^{-3}$ (details of this analysis, for one of the photos, is found in Fig. \ref{fig:density}). It is of course purely coincidental that the density estimates ended up this close together, but it still shows that bramblings can fly much denser that 0.04--0.1 m$^{-3}$. \citeA{Schifferli1953_OrnitBeob} made a similar exercise: he studied photos of a stream, taken from below, somewhat arbitrarily assumed a stream height of 4 m and counted 87 finches in 20 m$^2$. This results in a density of 1.1 m$^{-3}$. Having this said, based on observations at the Swedish roost, I would still say that bramblings not always fly this dense.

In general, I fully agree with Jenni that (i) number estimates via flock volume and density should be avoided, especially if these quantities are not carefully measured, and (ii) number estimates are much better done by estimating the intensity of birds passing an observer, preferably with the help of photographs.

So, how many birds could there have been in Thun? Here it is worth mentioning that there are other assessors than Jenni that has proposed lower numbers than M\"{u}hlethaler. \citeA{Egli1951_VogelHeimat}, for example, states that it was impossible to make any proper count but that 5 million should not be an overestimation. Moreover, according to \citeA{JenniNeuschultz1985_OrnitBeob}, a certain Frutiger reached the number 1--2 million after analysis of photos. That numbers disagree is one thing -- they often seem to do in the context of large flocks -- but the question is who made the least inaccurate estimate. It should be noted that the lower numbers presented are also rather poorly founded.  There is, however, yet another aspect to consider: the roost area. M{\"u}hlethaler specified the area to in total 13.6 ha, and that it comprised trees of Europan spruce \emph{Picea abies} and European silver fir \emph{Abies alba} at the age of 20--50 years. Compared to other roosts, many carefully studied, the resulting density of 5.3 million/ha appears surprisingly large. Although impossible to prove, I suspect that an overestimation has been caused by overestimation of both stream width and density. That the stream was 200 m wide and as dense as 1 m$^{-3}$ continuously for 45 min sounds rather extreme (around 13 000 bird per second, for each roost).

\subsection{Regarding the claim of 120 million}
As can be seen in Table \ref{tab:roost_table}, it is rather common with number estimates based on flock volume and density (method B). In fact, the 72 million in Thun is not the highest number reached in this manner. \citeA{NardinBrauchle1979} concluded that the roost in Etobon 1977--1978 comprised 100--120 million bramblings. Their analysis is somewhat difficult to follow, even for a physicist, but their argument behind setting the flock width and height to 250 m and 50 m, respectively, is not convincing. It is interesting to note that, along with this volume/density calculation (method B), they also present an intensity-analysis along the lines of method A. There, an analysis of a photo resulted in a total number of 12 million. 12 million seems reasonable for this 16 ha roost, and it can be noted that \citeA{Francois1978_Falco} in his description on this roost mentions, without any motivation however, the number 10 million. It can perhaps have been more birds than that, but based on the discussion above, I agree with \citeA{JenniNeuschultz1985_OrnitBeob} that the range 100--120 million is unrealistic. This may be an example of how previous overestimations drive new overestimations. In fact, \citeA{NardinBrauchle1979} explicitly states that they consider their analysis more restrained than \citeA{Muehlethaler1952_OrnitBeob}, arguing that if they would use his density of 1 birds/m$^3$ their estimate would increase to 875 million. While they may have been more careful in terms of bird density assumption, their values for the height of their stream, which they set to 50 m, is less conservative. Again, in my opinion, an example of how difficult it is to set flock dimensions correctly.

\subsection{Annual mass concentrations in corn districts in France 1960--1980}\label{subsec:corn_districts}
In previous sections it has been argued that the largest published values for brambling roosts should be regarded as overestimations. Clearly, it is not easy to answer the question of how many birds that the largest bramblings roosts has contained. The largest number mentioned in the review by \citeA{Jenni1987_OrnisScand} is 20 million from a roost in Pau, Pyr{\'e}n{\'e}es-Atlantiques in France. The roost in Pau is not just another large roost -- it represents fascinating brambling history. Corn production increased in France during the twentieth century and bramblings adapted to this new and abundant source of food . The roost in Pau  was likely established in early 1960s and was occupied annually for several years \cite{Alberny1965_OiseauxDeFrance}. This is itself an interesting contrast to other very large roosts which are formed during beech mast years and therefore used in isolated years (never several years in a row). This interesting adaption to changes in human farming was studied H{\'e}mery and Le Toquin, focussing on the energy needs of bramblings and the supply offered by the losses in corn farming \cite{Hemery1975_CrAcadSc,Hemery1976_LeTerre}. Changes in corn farming around 1980 (early burial of stubble) deprived the bramblings of this abundant food resources \cite{Dubois2008_BirdsFrance}.

Regarding the specific value of 20 million, the most common source for this value is an article on the energy expenditure of bramblings by Georges H{\'e}mery and Alain Le Toquin \cite{Hemery1975_CrAcadSc}. For example, the reference work \emph{Nouvel inventaire des oiseaux de France} by \citeA{Dubois2008_BirdsFrance} also refers to this work when stating 20 million as the peak of the roost numbers in Pau\footnote{Note that the reference given by \citeA{Dubois2008_BirdsFrance} is incorrect: the cited article is titled \emph{Determinisme energetique des concentrations de Pinsons du Nord\ldots}, not \emph{Determinisme des concentrations de Pinsons du Nord\ldots}. The missing word \emph{energetique} constitutes an unfortunate typo, as it indicates that the focus is on determining concentration rather than energy expenditure. The typo may simply come from the winter atlas of France, as this work contains the very same typo \cite{Hemery1991_WinterAtlas}}. Indeed, H{\'e}mery's article contain a datapoint in a scatterplot that corresponds to a roost number of 20 million. The value is, however, not commented further. According to \citeA{Hemery1976_LeTerre}, the number 20 million refers to the number in December 1967, but I have not found any work in which the number estimate is elaborated. Sadly, the ornithologist and brambling enthusiast Georges H{\'e}mery, passed away in 2013 \cite{Yesou2014_RevEcol} and could not give his view on these numbers. His colleauge Alain Le Toquin believes that the value 20 million comes from Jean-Claude Alberny (Le Toquin 2020, personal communication). However, Jean-Claude Alberny does not recall any other number than 15 million, a number he published in 1965 in an intriguing article describing the roost in Pau \cite{Alberny1965_OiseauxDeFrance}. The roost, in 1965 covering around 10--13 ha, was shared with starlings \emph{Sturnus vulgaris}. While most roosts has been located to conifers such as European spruce \emph{Picea abies}, European silver fir \emph{Abies alba} or Lawson cypresses \emph{Chamaecyparis lawsoniana}, birds here perched in Holly \emph{Ilex aquifolium}, an evergreen, and stunted oaks \emph{Quercus} with retained dead leaves (marcescence). The bramblings were popular among hunters (not only in Pau) and served as food, and ringing revealed that many bird were injured:

\begin{quote}
Le baguage a appris \'{e}galement l'important pourcentage d'oiseaux bless\'{e}s, pr\`{e}s de 5\% en janvier, chiffre diminuant par la suite. En effet les chasseurs appr\'{e}cient cette esp\`{e}ce, aussi bien dans les Landes, le Gers et les Basses-Pyr\'{e}n\'{e}es o\`{u} l'on peut trouver des brochettes enti\`{e}res de ces "ortolans".
\end{quote}

Alberny also reported that some birds suffered from some kind of disease. Regarding the number of birds Alberny suggests 15 million, as also stated in \ref{tab:roost_table}. This is, however, a rough estimate. Birds came from all directions and proper counting seemed impossible. Alberny, believing it to be an underestimation, writes:

\begin{quote}
Une é\'{e}valuation parait impossible, les oiseaux arrivant de tous c\^{ô}té\'{e}s à la fois. De plus la configuration du relief ne permet pas de voir le dortoir en entier. Les Pinsons sont de tr\`{e}s loin plus nombreux que les Etourneaux. On peut affirmer sans crainte qu'il y en a au moins 15 MILLIONS... Probablement plus...
\end{quote}

It seems unlikely that H{\'e}mery studied the numbers in Pau as early as 1967. In an ambitious article from 1981, he elaborates on the difficulties of counting large numbers of birds, and explores whether infrared thermography can be of assistance \cite{Hemery1981_Thermographie}. However, in this work H{\'e}mery writes that proper counting at a roost requires multiple competent observers and argues that uncertainties may be as high as 50\% when dealing with gatherings of birds on the order of ten millions. H{\'e}mery also presents the results of two visual counts made in Pau in February 1979: $2.3$ million $\pm 25\%$ on February 5 and later, when most birds had abandoned the roost, $450\,000 \pm 33\%$ on February 14. The work does not contain any comparison with earlier numbers from Pau, nor details on how and by whom the counting was conducted in 1967 when 20 million was reported.

To conclude, it is difficult to confirm that there was as much as 20 million in a single roost in Pau. Given its large area (9--13 ha), the roost in Pau is without doubt one of the largest ever registered. As can be seen it Table \ref{tab:roost_table}, there are a few roost that has been of similar size. Which one of these that held the highest number of bramblings remains unknown.

\subsection{The classic 1915--1916 roost in Sweden}
Another interesting example is the internationally renowned winter roost in K{\aa}ger\"{o}d, Sweden 1915--1916, studied in detail by Hugo Granvik \cite{Granvik1916_FaunaFlora,Granvik1916_JOrnit,Nilsson1983_Anser}. In an otherwise detailed and careful study of a brambling roost, the number 5.4 million given by Granvik lacks detailed justification. Granvik states 5.4 million is reached if assuming that 2 000 birds per second arrive at the roost during the observed 45-minute fly-in duration. The number in itself is not sensational, given what we know today, but as the roost area appears to have been rather small the number appears surprisingly high. Granvik specifies that roost area to "n{\aa}got tunnland" in Swedish, literally meaning around one barrel of land, i.e. 0.5 ha. That the roost area was rather small is also supported by sketches made by \citeA{Nilsson1983_Anser}. It cannot, of course, be ruled out that the accumulation of bramblings at this site as the winter progressed -- in relation to available area -- made this roost extraordinary dense. In fact, \citeA{Granvik1916_FaunaFlora} writes that the bramblings utilized birch-trees in the border of the coniferous roost area. It is however not clear whether it was confirmed that these trees were populated also in the middle of the night. In the 2019--2020 roost in Sweden, I noted movements from beech trees into the actual roost also after dusk.

\subsection{Counting trees and birds per tree}
The study by \citeA{Kjellen1993_Anser} is, to the best of my knowledge, the only previous work that have conducted number estimation by estimating the number of trees and birds per tree (method C). The work concerns a roost in Sweden located in January 1993, comprising approximately 7.5 ha (500 m by 150 m) of plantations of European spruce of different ages and heights between 10 and 20 m. Furthermore, it was estimated that there on average was one tree per 10 m$^2$, resulting in a total tree count on 7 500. The authors estimated that at least 200 birds sat in each tree, and perhaps 500 in the larger ones, and conclude that the total number of bramblings at the very least should be 2 million. A fly-in count (method B, but without photos) by \citeA{Lithner1995_FaglarNV} also estimated the number of birds to at least 2 million. Since the stated roost area is rather large, the number may appear somewhat low. It is, however, worth noting that this particular roost seems to have suffered from some kind of deadly disease (large numbers of dead birds were found within the roost area).
As described in Section \ref{sec:the_mass_conc_in_Sweden}, observations at the Swedish 2019--2020 roost suggest that a single tree can host on the order of 1000 finches. A similar number was suggested for larger trees in the well-studied roost in Switzerland 1916--1947 \cite{Gueniat1948_OrnitBeob}. More studies on how many birds that roost per tree would be enlightening.

\subsection{High numbers from nowhere}
Around 100 texts on bramblings were read as a part of this review and on a few occasions I came across very high numbers that I could not track: the cited articles did not, as far as I can see, contain the claimed information. Unless other researchers have better luck in finding support in original sources, these numbers should not be recited. To facilitate future works on the topic, I have decided to explicity mention these oddities in the vast literature on brambling mass concentrations.
The perhaps most important case is a claim in the standard reference \emph{Finches} by \citeA{Newton1972_Finches}. Referring to \citeA{Gueniat1948_OrnitBeob} and \citeA{Sutter1948_OrnitBeob}, Newton writes that there was a roost of 50 million bramblings in Switzerland during the 1946--1947 winter. None of these articles contain such a number (perhaps a referral in \citeA{Gueniat1948_OrnitBeob} to an older roost and the number 5$\nicefrac{1}{2}$ million was misinterpreted as 51--52 million and rounded to 50 million?). Referring to Newtons text, this numbers is unfortunately recited in other important works. Another large number that lacks support is the 61 million that \citeA{Moller2019_EcoEvol} recites, referring to \citeA{Geroudet1952_NosOiseaux}. Finally, it should be noted that \citeA{Chalverat2003_SchwForst} incorrectly refers to the PhD thesis of Jenni \cite{Jenni1984_PhDThesis} when stating that the roost in R\"{o}serental 1977--1978 comprised 28 million birds. Instead, the thesis argues (p. 40--41) that the roost contained around 6 million birds (the range 2--9 million is also given). This estimation has also been published in journals \cite{JenniNeuschultz1985_OrnitBeob,Jenni1991_OrnisScand}

\section{Conclusions and discussion}
The literature review presented in this work suggest that there is support for bramblings roost involving up to around 15 million bramblings. In terms of areal density of birds, a roost may hold on the order of one million birds/ha. As large variations in roost density can be expected, this should be seen as an utterly rough estimate in need of further verification. As a comparison, the passenger pigeon with its tenfold mass has been estimated to roost in densitites of around 100 000 birds/ha \cite{Ellsworth2003_ConsBiol} and the red-billed quelea has been claimed to roost at densities of 2.5 million/ha \cite{Manikowski1988_TropicalPest}. Claims of higher numbers, or densitites, such as the 70 million bramblings from Switzerland 1951--1952, is based on questionable methods and does not fit well the overall picture that emerges from collection of reports on mass concentrations of bramblings that has been built up throughout the years. Unfortunately, despite that the number 70 million has been rejected earlier \cite{JenniNeuschultz1985_OrnitBeob}, it keeps being recited without reservations in new works on mass appearances of birds. A main reason for this is perhaps that important reference books, like \emph{Finches} by \citeA{Newton1972_Finches} and \emph{Bird Migration} by \citeA{Alerstam1993_BirdMigration}, presents these values without reservations or discussions of uncertainties (in later works such as \emph{Population Limitation in Birds}, however, \citeA{Newton1998_PopLim} adheres to Jenni’s more restrictive view that values above 20 million lack proper support).

Given the difficulty in counting large numbers of birds, reports of very large numbers should always be used and recited with great care, regardless of whether it concerns bramblings, starlings, blackbirds, queleas or passenger pigeons. The large ecological and economical significance of large flocks of birds motivates further studies. Clearly, much can be done to improve our knowledge on the number in which flocks of our most abundant birds appear. This includes bramblings, despite all the efforts made so far. For example, there is no publication with a solid analysis and presentation of uncertainties in the number estimation conducted. In addition, roost areas are often not well investigated (demarcation in map missing). Just as for example vagrancy and taxonomy gets a lot of attention in ornithological journals, it would be very valuable to see also more work on accurate estimation of flock and roost sizes. From a critical point of view, \emph{millions} should only be interpreted as \emph{many} unless carefully motivated. While H{\'e}mery argued that, even with multiple competent observers, uncertainties may be 50\% when numbers are in the ten million range \cite{Hemery1981_Thermographie}, the discussion on the 70 million and 120 million numbers \cite{Muehlethaler1952_OrnitBeob,NardinBrauchle1979,JenniNeuschultz1985_OrnitBeob} indicates that uncertainties also can be a factor ten, depending on count method.

\subsection{Roost characterization: a short checklist}
\begin{itemize}
  \item \textbf{Counting.} Try to monitor morning lift or evening fly-in from a spot with good overview. Make careful notes on timing (start and end of movements) and estimate the intensity of birds streams preferably by taking systematic series of photographs (or film). Avoid estimating numbers from flock volumes (width, height and length) and bird densities (birds per m$^3$).
  \item \textbf{Roost area.} Determine roost boundaries via studies of excrement layers. Although zones around the roost, where birds gather before flying in, will show a lot of droppings, the actual roost area will be completely covered in excrements. Use the demarcation to estimate the roost area carefully using map tool (i.e. Google Maps).
  \item \textbf{Document trees.} Document trees in the roost (species, age, height and distance between trees) and try to estimate the total number of trees.
  \item \textbf{Birds per tree.} Try to estimate the number of birds per tree, preferably during the night using thermal cameras.
  \item \textbf{Roosting in bare trees.} If there are indications that bird sleeps also in non-conferious trees (e.g. in beech or other deciduous trees), confirm this with e.g. a thermal (infrared) camera in the middle of the night. The bramblings move also after dusk, and evidence of roosting in, for example, peripheral deciduous trees would be interesting (cf. \citeA{Granvik1916_FaunaFlora}).
\end{itemize}

\section{Acknowledgement}
First of all, I would like to thank Hanna, Siri och Vera for letting our family include also bramblings for quite some months. There are also many that have contributed more directly to the work, for example by engaging in interesting discussions, supplying complementary information, assisting in my hunt for obscure references, lending me equipment or reviewing and providing feedback on the manuscript. In particular, I would like to acknowledge the help from Thomas Alerstam, Johnny J{\"o}nsson, Nils Kjell\'{e}n, {\AA}ke Lindstr\"{o}m, Cecilia Nilsson, Lars Rippe \& Roine Strandberg (Lunds Universitet), Sara Agrup, Roland \& Jean-Claude Alberny, Janne Dahl\'{e}n (Ecogain), Igor Drobyshev \& Rolf {\"O}vergaard (SLU), Miro Fulín, Mats Hellmark (Sveriges Natur), Nicole H{\'e}mery, Alain Le Touqin, Lukas Jenni \& Hans Schmidt (Schweizerische Vogelwarte, Schweiz), Toma\v{z} Miheli\v{c} (Birdlife Slovenia), Yves Muller (Ligue pour la protection des oiseaux, LPO, Frankrike), Karin Persson (Falsterbo F{\aa}gelstation), Jean-Marc Pons \& Jean-Pierre Siblet (Mus{\'e}um national d'Histoire naturelle, Paris), Fritz Rosen\"{o}rn (R\"{o}ssj\"{o}holms gods), Sissel Sj\"{o}berg (University of Copenhagen), my brother Stefan Svensson, Jean-Marc Thiollay, Raffael Winkler (Naturhistorisches Museum Basel, Schweiz), Hasti Yavari (Axis Communications), Pierre Y{\'e}sou and Jabi Zabala.

\bibliographystyle{apacite}
\bibliography{Brambling_references}

\end{document}